\documentclass[12pt]{article}

\usepackage[latin1]{inputenc}
\usepackage[english]{babel}
\usepackage{amssymb,amsmath}
\usepackage{amsthm}
\usepackage{amsfonts}
\usepackage{mathtools}
\usepackage{graphicx}
\usepackage[margin=2.0cm]{geometry}
\usepackage{enumerate}
\usepackage{color}
\usepackage[noadjust]{cite}
\usepackage{tikz}
\usepackage{mathtools}
\usepackage{framed}
\usepackage{subfig}
\usepackage{booktabs}
\usepackage{hyperref}
\usepackage{array}
\usepackage{color, colortbl}
\usepackage{multirow}

\theoremstyle{definition}

\newtheorem*{conj*}{Conjecture}

{
\begin{framed}\sffamily \textbf{#1}%
}%
{\rmfamily\end{framed}}

\usetikzlibrary{shapes, graphs, graphs.standard, calc}
\tikzstyle{vertex}=[circle, draw, fill=black, minimum size=8pt, inner sep=0pt]

\makeatletter  

\title{Addressing the cold start problem in privacy preserving content-based recommender systems using hypercube~graphs}

\author{
Noa Tuval\textsuperscript{1}
Alain Hertz\textsuperscript{2},
Tsvi Kuflik\textsuperscript{1},
\\[3mm]
\footnotesize \textsuperscript{1} Information Systems Department\\
\footnotesize The University of Haifa\\
\footnotesize \textsuperscript{2} Department of Mathematics and Industrial
Engineering\\
\footnotesize Polytechnique Montr\'eal - Gerad, Montr\'eal, Canada\\
\footnotesize Corresponding author. Email: alain.hertz@gerad.ca\\[3mm]
}

\DeclarePairedDelimiter \round{\lfloor}{\rceil}

\begin{document}

\maketitle
\vspace*{0.2cm}

\hrule
\vspace*{0.2cm}
\small
\noindent
\textbf{Abstract.} \\
The initial interaction of a user with a recommender system is problematic because, in such a so-called cold start situation, the recommender system has very little information about the user, if any. Moreover, in collaborative filtering, users need to share their preferences with the service provider by rating items while in content-based filtering there is no need for such information sharing. We have recently shown that a content-based model that uses  hypercube graphs can determine user preferences with a very limited number of ratings while better preserving user privacy. In this paper, we confirm these findings on the basis of experiments with more than 1,000 users in the restaurant and movie domains.  We show that the proposed method outperforms standard machine learning algorithms when the number of available ratings is at most 10, which often happens, and is competitive with larger training sets. In addition, training is simple and does not require large computational efforts.

\vspace*{0.2cm}

\vspace*{0.2cm}
\hrule

\normalsize

\section{Introduction} \label{sec_intro}

The growing involvement of recommender systems in our daily life raises the need to develop reasoning mechanisms to recommend relevant items to users based on their past interactions with the system as well as additional external sources that may be available. In order to personalize the recommendations, these systems create a model for each user on the basis of information previously collected 
\cite{Aggarwal16}.
 The two prevailing classical user modeling methods are {\em collaborative filtering} (CF) and  {\em content-based filtering} (CB). 
 Recommender systems based on CF build such models based on the similarities of preferences between users: the items recommended for a user are those that received a high score from a group of users who show similar ratings. Conversely, items recommended to a user by CB systems have attributes similar to those of items that have received a high rating from the user in the past.

CF systems need a lot of data to infer the preferences of their users and have difficulty in producing good recommendations when there are very few users compared to the number of items. Moreover, CF systems suffer from {\em the new item problem}: new products with no rating history are unlikely to be recommended \cite{Ricci2022,Zhu2021}. This problem does not exist in CB systems because recommending an item depends solely on its attributes. 
On the other hand, CB methods have problems that do not exist in CF systems: describing the items using a limited number of attributes makes it difficult to represent all aspects of the items, which may impair the accuracy of the recommendations. In addition, CB systems have the problem of over-specialization since they only recommend items with similar attributes to items the user has liked in the past \cite{Aggarwal16}.

With the huge amount of data flowing in the network, awareness of user privacy is increasing, due to the sensitivity and vulnerability of user data \cite{Himeur22}. Recommender systems require users to reveal their preferences, in order for the recommendations produced by the system to match their tastes \cite{Aggarwal16}. However, most users want the system to maintain their privacy \cite{Gunawardana2022}. Since CF methods rely on data provided by a group of users to generate recommendations for a particular user, this approach requires many users to be identified and share their preferences with the system in the form of ratings \cite{Aggarwal16,Gunawardana2022,Koren2022}. Conversely, CB methods need to know the characteristics of each item and user preferences about these characteristics to recommend to users items that match their tastes \cite{Burke2007,Aggarwal16}. These preferences can be stored locally on the user's personal device  and not shared \cite{Kuflik2009b}. In this aspect, CB systems preserve the privacy of users better than CF systems. 

An improvement to the classic recommendation methods are the {\em hybrid} methods \cite{Ricci2022,Cano17} and {\em graph-based} methods \cite{Aggarwal16,Wang21,Ricci2022}.
Hybrid recommender systems combine different methods, where usually the main method used is CF in combination with another strategy, while exploiting the advantages of the participating methods to overcome some difficulties such as the new item problem and data sparsity \cite{Cano17,Ricci2022}. One of the challenges facing hybrid systems is the need for data from different sources in order to create good personalized recommendations. However, in addition to the additional space required, the use of large amounts of data also requires more intensive calculations. 

Graph-based methods provide versatile structures for representing the relationships among users and items \cite{Aggarwal16,Minkov,Tiroshi}. In recent years, {\em graph learning} (GL) approaches have been developed for graph-based user modeling. Unlike the CB and CF methods, GL techniques can extract knowledge from various graph representations even where the entities are implicitly connected \cite{Wang21,Ricci2022}. A major challenge of graph-based recommender systems is that the data is usually very large and requires a lot of time and complex algorithms to be processed
 \cite{Wang21}.

It is  natural to question what happens when a recommender system does not have enough information to define an adequate model for each user. This problem, which prevents new users from taking full advantage of the power and relevance of recommender systems, is known as the  {\em cold start problem}. It prevents CF methods from producing reliable recommendations since there is no (or too little) prior information about the user of the system which therefore cannot determine users with similar tastes to produce recommendations \cite{Vinagre}. Moreover, since CF systems derive user preferences based on previous ratings, they must have enough ratings to create reliable user models \cite{Ado,Musto2022}.

\begin{figure}[!htb]
		\centering
		\includegraphics[scale=0.74]{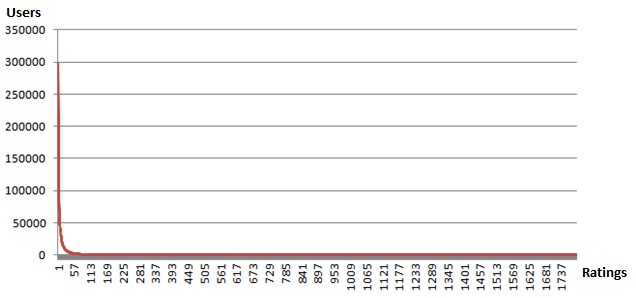}
		\caption{The long tail distribution of the number of ratings per user (taken from the yelp website https://www.yelp.com/dataset).}
		\label{fig:Figure1}
	\end{figure}

As stated in \cite{Sinha,Zhang22}, the cold start problem is one of the research challenges of recommender systems. The problem has caught the attention of many researchers \cite{Bernardi,Gupta} because the ability of a recommender system to produce relevant recommendations often comes up against the long-tail distribution of the number of ratings per user. As an example, we show in Figure \ref{fig:Figure1} the number of ratings that the users of the yelp website (https://www.yelp.com/dataset) have given. It is clear that the vast majority of users of the system have provided very few evaluations. For example, there are 1,069,244 users with less than 10 ratings, but only 13,351 users with at least 20 ratings.

We have recently described a new approach to create user models with CB systems \cite{HKT21}. Roughly speaking, the users of the system and the items that can be recommended to them are vertices of a hypercube $Q_n$ of dimension $n$ (see Figure \ref{fig:Figure00} in Section \ref{sec:3} for an example with $n=3$), where $n$ is the number of attributes of the items. Given a set $R$ of items evaluated by a user, the estimated user model is the vertex $u$ of the hypercube such that the distances between $u$ and the vertices of $R$ correspond as closely as possible to the scores assigned by the user to these items. Preliminary results reported in \cite{HKT21} show that this estimate is generally very accurate when using it to predict user preferences for items. 
In this paper, we confirm these observations on the basis of larger scale experiments. 

\section{Objectives and practical implications}

Our main objective is to demonstrate that the content-based recommender system based on hypercube graphs described in \cite{HKT21} is an interesting approach to solve the cold start problem. We will show that it outperforms standard machine learning algorithms when the number of available ratings is at most 10.
Also, an advantage of our approch over CF methods is that it allows preserving the confidential data of users because the information necessary to identify the preferences of each user can be kept locally on a laptop or a smartphone.

In summary, we show that we can learn user preferences with very few ratings. The comparison between the results obtained by our method with those produced by current standard techniques confirms the ability of our approach to solve the cold start problem while preserving user privacy.

The paper is structured as follows: A brief literature review related to user modeling, to the cold start problem and to graph-based models for recommender systems is given in the next section. Our approach based on hypercube graphs is described in Section \ref{sec:3}, and computational experiments are reported in Section \ref{sec:4} where we compare our algorithms with five commonly used machine learning techniques. Concluding remarks and future work are given in Section \ref{sec:5}.

\section{Background and related work} \label{sec_prop}

We start this section with an overview of models and techniques used in recommender systems, and standard techniques used to build such models. We then show how graph theory is used in such systems, and we conclude the section with a review of recent studies for solving the cold start problem in recommender systems.

\subsection{User models and machine learning techniques}

Efficient user modeling is a major challenge in recommender systems that aim at providing personalized service to their users. These systems leverage information collected about users from various sources such as web pages, social networks, and e-commerce sites. The information collected can then be structured in different ways, depending on the chosen recommendation technique \cite{Ricci2022}.

Recommender systems usually make use of either or both collaborative filtering (CF) and content-based filtering (CB) approaches, as well as other techniques such as community-based methods and knowledge-based methods \cite{Ricci2022}. CF approaches recommend to the user items that users with similar taste like. These methods often use latent factor models like matrix factorization methods to improve data representation by transforming users and items to the same latent factor space \cite{Aggarwal16,Koren2022}. Other methods used by CF are various machine learning techniques which build a summarized model of the data and predict items that the user may have an interest in \cite{Portugal}. 
CB approaches utilize user preferences over a series of item attributes to recommend additional items with similar properties \cite{Ricci2022,Musto2022}. CF and CB recommender systems need a sufficient amount of previous user ratings to generate accurate recommendations. However, these systems are not suitable for highly customized products which are not purchased on a regular basis, such as houses, cars or even smart electronic products \cite{Aggarwal16}. Moreover, these systems tend to recommend items that are most similar to the ones they users have already selected \cite{Aggarwal16}. 

Current recommender systems may also combine one or more approaches into a hybrid system which tries to exploit the benefits of each of the combined methods \cite{Burke2007,Ricci2022}. One of the hybrid techniques is the cross-representation mediation of user models from CF to CB recommender systems \cite{Berkovsky08,Berkovsky09}. This mediation technique is effective in cases where the lack of data in a CB recommender system makes it difficult to generate recommendations. Using the mediation process, the CB system can leverage the data of the same users that has been collected by a CF system and transform these data into weighted item attributes.

In recent years, modeling user behavior using classical machine learning techniques has become a challenge with the large amount of information available for recommender systems and the complexity of user interactions with the system \cite{Zhang22}. Thus, deep learning techniques have been increasingly used in recommender systems as they enable processing unstructured data and inferring hidden patterns of user preferences and item representation \cite{Zhang22,Musto2022}. 
Also, deep learning techniques are able to consider descriptive information (such as text, images, audio, and video) about users and items available from various sources, to create a more reliable and accurate user model \cite{Zhang22}. Deep learning has been successfully applied in fields as diverse as movie, music, image or quote recommendation. \cite{Guan19,Tan16,Oord13,Niu18}.

As the architecture of a neural network has a decisive influence on the learning model performance, improving the construction of the neural system, for example by adding hidden layers to the neural network, may help the multitask learning. Also, implementing recommendation algorithms through deep neural networks involves high computational cost that should be reduced using more efficient algorithms \cite{Zhang22,Dau20}. The big amount of data required for the learning process is another disadvantage of deep learning techniques, since it is not possible to produce accurate recommendations when there is not enough data available \cite{Zhang22}.

\subsection{Graph-Based User Modeling}

Graphs and network are increasingly applied for user modeling. The main advantage of these models is that they allow combining entities and their mutual links in simple structures, without loss of information \cite{Tiroshi}. At the same time, graph learning methods, such as random walk methods, graph neural networks methods and graph embedding methods, capable of learning complex relations between entities and extracting knowledge from graphs, were developed \cite{Wang21}.

In most graph-based approaches, user and item entities are represented by vertices, and an edge connects a user $u$ with an item $i$ when u has rated $i$, which leads to a bipartite graph \cite{Zhang22}.
An edge connecting a user to an item he has evaluated can be weighted with the value of the corresponding score, and the amount of information that can propagate along such an edge is then proportional to this weight \cite{Nik22}. Graph models are used to detect similarities between users or items, these similarities being measured using random walks \cite{Jackson} or path-based approaches \cite{Nik22}.

The bipartite graph described above can be extended by adding additional attributes as new graph vertices and linking them to the existing vertices \cite{Wang21}. The resulting graph is called {\em heterogeneous} and enables integrating collaborative ratings with domain knowledge and content aspects for better personalized recommendations \cite{Minkov,Wang21}. 
User modeling with heterogeneous graphs is used, for example, in \cite{Wasserman} for movie recommendation and in \cite{Minkov} for on-site guidance of museum visitors.

Graph Neural Networks (GNN) is a deep learning approach designed to help understanding all graph components and hidden data, which is essential for recommender systems.
GNN techniques have proved their potential to learn from graph structured data, for example for session-based and social recommendation \cite{Qiu20,Fan19}. However, a major challenge of this model is the low computation efficiency due to the big amount of data required for the learning process \cite{Zhang22}.

A new approach for efficient analysis of graph structured data is graph embedding which converts graph data into low dimension vectors while preserving the graph structure and connectivity \cite{Cai18}. For example, an edge embedding technique is proposed in \cite{Verma19} for friend recommendation and in \cite{Chen19} for time-aware smart object recommendation. 

To conclude this section, let us mention that graphs are also used to produce explainable recommendations, which is an aspect of recommender systems that has received recent attention \cite{Ma19}.

\subsection{The Cold Start Problem in Recommender Systems}

When someone starts using a recommender system, or when his interaction with it is superficial, there is not enough information available for the system to create a reliable model of the user. This situation where the system has not yet gathered sufficient information to understand the user preferences is known as the cold start problem. It refers to the situation where a user enters the system for the first time as well as to the situation where the user has little interaction with the system and therefore has only evaluated a few items. 

A common way to learn about user preferences and overcome the cold start problem is to ask users to rate a few items \cite{Rashid}. Based on this limited information, the system should create initial user models. It is therefore desirable that the items offered for evaluation can help the system to understand user preferences. 
Several techniques have been proposed for determining the most informative items to be rated by a user \cite{Rashid}. Recommending popular items is one of the solutions in cases where there is not enough prior information about user preferences. The obvious disadvantage of such a solution is that the user gets a recommendation for a product that has a high chance that he is already familiar with \cite{Aggarwal16}.

Sometimes users may wish to explore new items offered by a recommender system. In this case, they will cooperate with the system and willingly rate the suggested items. However,users may not have the time or desire to rate a large number of items. They want to get recommendations without being too involved in the process. Requiring users to rate items as a precondition for creating recommendations for them can be considered intrusive. It would therefore be interesting to be able to measure the benefit of getting additional ratings in terms of the increased accuracy of predictions based on these ratings \cite{Ado}.

Another way proposed to address the cold start problem is to use hybrid recommender systems that combine different types of recommendation techniques in order to overcome the weakness of one component through the use of another component of the system. A hybrid approach is described in \cite{Berkovsky08} that uses a modeling mediation process which integrates partial user models produced by different types of recommender systems. When a new user enters a CB recommender system that has no information about his preferences, this cold start situation is overcome through a mediation process that leverages previous user data collected in a CF system in the same domain. The user model produced by the CF method, which is represented by a vector of ratings, is converted in the mediation process into a vector of weighted attributes which is then used by the CB system to produce recommendations \cite{Berkovsky08}.

A cross-domain recommender system is proposed in  \cite{Wang20} where data from an online shopping domain is combined with information taken from an advertising platform.  A neural CF model is trained on users who appear in both domains, allowing the system to produce recommendations to users who are active on the advertisement platform but not in the online shopping domain. 

A deep learning CF approach that is suitable for both warm start and cold start is proposed in \cite{Vol17}. This method considers the cold start problem as a problem of missing data, so that the training of the model is done via input dropout, i.e., some nodes along with their connections are dropped from the neural
network during training to estimate how each rating contributes to improving the accuracy of user models. 

Few-shot learning (FSL) is a machine learning method that aims to classify new data when only a few training samples are available. It has been used to solve the cold start problem in recommender systems \cite{Hao20,Wang20}. 
Zero-shot learning (ZSL) is a variant of FSL that tries to deal with the situation where a learner observes samples from classes which were not observed during training, and needs to predict the class that they belong to. A method for solving the cold start problem from
a ZSL perspective is proposed in \cite{Li19}.
Also, collaborative filtering algorithms based on meta-learning, such as MAML (model-agnostic meta-learning) help the recommender system to initialize new users, 
while overcoming the bias that may be caused by  global sharing initialization parameters common to all users
\cite{Lee19,Dong20}. All the above techniques are based on learning many user profiles based on their personal data. For comparison, the method studied in this article
handles the cold start issue without requiring users' personal data.

In the next section, we describe a technique we proposed in \cite{HKT21} for building user models through hypercube graphs in a CB recommender system. It differs greatly from approaches based on deep learning and graph embedding. 

\section{Hypercube graphs for CB recommender systems}\label{sec:3}


In this section we briefly provide a theoretical background about using hypercube graphs for CB representation of items and users, and we then describe two user modeling techniques proposed in \cite{HKT21} based on hypercube graph representations in CB recommender systems.

\begin{figure} [h!]
	\begin{center}
		\includegraphics[scale=0.20]{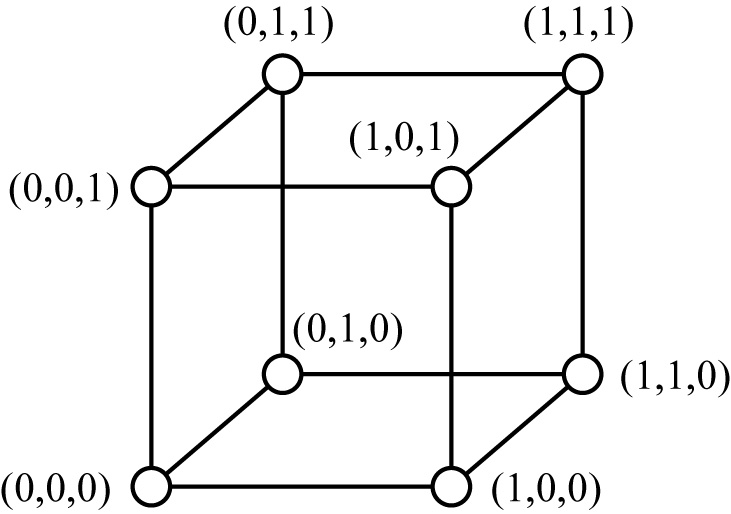}
		\vspace{-0.2cm}		\caption{The hypercube $Q_3$.}
		\label{fig:Figure00}
	\end{center}
\end{figure}

Let $U$ be a set of users of a recommender system, let $A$ be an ordered set of $n$ Boolean attributes, and let $I$ be a set of items.
Let $Q_n$ be the $n$-dimensional hypercube with vertex set $\{0,1\}^n$, and where two vertices $\boldsymbol{x}$ and $\boldsymbol{y}$ are linked with an edge if and only if the Hamming distance $d(\boldsymbol{x},\boldsymbol{y})$ between $\boldsymbol{x}$ and $\boldsymbol{y}$ (i.e., the number of indices $i\in \{1,\ldots,n\}$ such that $x_i\neq y_i$) equals 1. Items and user models are represented as vertices in $Q_n$ (see Figure \ref{fig:Figure00}). More precisely:
\begin{itemize}
	\item A vertex $\boldsymbol{v}^i=(v^i_1,\ldots,v^i_n)$ of $Q_n$ is associated with every item $i\in I$ so that $v^i_j=1$ if $i$ has the $j$th attribute in $A$, and $v^i_j=0$ otherwise;
	\item a vertex $\boldsymbol{w}^u=(w_1^u,\ldots,w_n^u)$ of $Q_n$ is associated with every user $u\in U$ so that $w_j^u=1$ if and only if $u$ `likes' the  $j$th attribute in $A$.
\end{itemize}

Note that two items with the same attributes and two users with the same preferences are associated with the same vertex in $Q_n$. We can therefore consider every vertex of $Q_n$ as an {\it item type} and a {\it user type}.

Assume that a user $u\in U$ of a recommender system has indicated the number of attributes $a\in A$ he does not like in every items of a subset $I'\subseteq I$ of rated items.
This means that we know the Hamming distance between $\boldsymbol{w}^{u}$ and all vertices $\boldsymbol{v}^i$ with $i\in I'$. To identify the vertex $\boldsymbol{w}^{u} $ which corresponds to the model of $u$, it is therefore sufficient to determine a vertex $\boldsymbol{x}$ of the hypercube such that, for all $i\in I'$, the Hamming distance $d(\boldsymbol{x},\boldsymbol{v}^i)$ is equal to the
number of attributes that user $u$ does not like in item $i$.

The problem we are facing with this approach is that users never indicate this precise information (i.e. the Hamming distance) which would allow us to fully understand their tastes. Rather, they give a score to the items they rate, and we therefore have to convert this score into a Hamming distance in $Q_n$.  We suppose here that the users rate items according to an $s$-star scale, where the highest score of $s$ stars is given by users to items that perfectly match their preferences, and the lowest score of 1 star when they did not like any of the attribute values of the rated item. So let $r_i$ be the rating given by $u$ to an item $i\in I$, using an $s$-star scale. This rating can be translated into a distance $\delta_i$, called $d$-rating using function $\tau:\left\{1,\ldots,s\right\}\longrightarrow[0,n]$ which is defined as follows:

$$\delta_i=\tau\left(r_i\right)=n-\frac{n\left(r_i-1\right)}{s-1}.                 $$

The worst rating $r_i$ with 1 star is thus translated into distance $n$, and the best rating with $r_i=s$ stars into distance 0. The $d$-rating can be considered as an approximation of the Hamming distance. Indeed, if a user $u\in U$ likes all attributes of an item $i\in I$, his rating will be $r_i=5$ stars, which we translate into $\delta_i=0$, with the meaning that our estimate of the Hamming distance $d(\boldsymbol{w}^{u},\boldsymbol{v}^{i})$ is $\delta_i=0$. On the contrary, if $u$ does  not like any of the attributes of an item $i\in I$, his rating will be $r_i=1$ star, which we translate into $\delta_i=n$, with the meaning that our estimate of the Hamming distance $d(\boldsymbol{w}^{u},\boldsymbol{v}^{i})$ is $\delta_i=n$.

The models described in \cite{HKT21} for recommending items to users of the system are based on this concept. They aim to determine a vertex $\boldsymbol{x}$ that fits the ratings of a user $u\in U$. More precicely, assume that $u$ has rated a subset $I'\subseteq I$ of items, let $r_i$ be the rating assigned to $i\in I'$ by $u$, and let $\delta_1,\ldots,\delta_{|I'|}$ be the associated $d$-ratings. The task is to determine the vector $\boldsymbol{w}^{u}$ associated with $u$ in $Q_n$. Hence, the output $\boldsymbol{x}$ of the algorithms proposed in \cite{HKT21} should ideally be equal to $\boldsymbol{w}^{u}$. If $\boldsymbol{x}=\boldsymbol{w}^{u}$ then, for all $i\in I'$, the Hamming distance $d(\boldsymbol{v}^{i},\boldsymbol{x})$ should be the number of attributes $a\in A$ such that
$w^{u}_a\neq v^{i}_a$. Since we estimate  $d(\boldsymbol{v}^{i},\boldsymbol{w}^{u})$ as being equal to $\delta_i$, the cumulative error made by assuming that $\boldsymbol{x}=\boldsymbol{w}^{u}$ is estimated as $f_1(\boldsymbol{x})$, where
$$f_1(x)=\sum_{i\in I^\prime}|d(\boldsymbol{v}^i,\boldsymbol{x})-\delta_i|.$$
In summary, the first algorithm described in \cite{HKT21} generates a vector $\boldsymbol{x}$ in $Q_n$ with minimum value $f_1(\boldsymbol{x})$. 
Variations of this algorithm are proposed in \cite{HKT21}. We consider here the one where `don't care' values are allowed. In other words, it is assumed that it may happen that a user does not care about some item attributes. A user $u\in U$ is then represented by a vector $\boldsymbol{w}^u$ in $\{-1,0,1\}^n$ (instead of $\{0,1\}^n$), so that  $w^u_j=-1$ if $u$ does not like the $j$th attribute, $w^u_j=0$ if $u$ does not care about it, and $w^u_j=1$ if $u$ likes it.
If $w^u_j=0$, the ratings of $u$ do not depend on the value of the $j$th attribute. To take this into account, the Hamming distance $d$ is replaced by a new distance $d':\{0,1\}^n\times \{-1,0,1\}^n\rightarrow \{0,\ldots,n\}$ which, given a vertex $\boldsymbol{v}\in \{0,1\}^n$ and a vertex $\boldsymbol{x}\in \{-1,0,1\}^n$,
counts the number of components $j$ with $v_j=1$ and $x_j=-1$, or $v_j=0$ and $x_j=1$. The task of the second algorithm that we will test in the next section  is to determine a vertex $\boldsymbol{x}$ in $Q_n$ that minimizes $f_2(\boldsymbol{x})$, where $f_2$ is obtained from $f_1$ by replacing $d$ with $d'$.

As shown in \cite{HKT21}, determining $\boldsymbol{x}$ that minimizes $f_1(\boldsymbol{x})$ or $f_2(\boldsymbol{x})$ is a problem that can be formulated as an integer program. Preliminary experiments are reported in \cite{HKT21} which show that few ratings are sufficient to generate a vertex $\boldsymbol{x}$ which is close to the user model.
The aim of the next section is to confirm these findings on the basis of larger scale experiments.

\section{Computational experiments}\label{sec:4}

In this section, we report results obtained by executing the two algorithms mentioned in Section \ref{sec:3}. 
We  first focus on datasets from the restaurant domain which we used in \cite{HKT21} and compare the results produced by the two algorithms with those obtained using classical machine learning techniques widely used in recommender systems. We will then focus on datasets taken from the movie domain.

\subsection{Experiments with restaurant datasets}

In a previous study \cite{HKT21} we extracted ten datasets from the yelp website, each containing 500 restaurants that were rated by a single user on a 5-star scale. The number $n$ of attributes varies from 45 to 51, referring to the restaurant genre (Italian, Chinese etc.), the price (low, medium, high, very high), and some other facilities (reservations, outdoor seating etc.). A restaurant in the dataset corresponding to user $u_i$, is characterized by $n_i$ attributes from the full list of attributes, where an attribute is included in the dataset only if it exists in at least one of the restaurants rated by the user. For example, since no restaurant that user $u_7$ has rated serves Greek/Turkish food, the attribute corresponding to this type of food does not appear in $u_7$'s dataset and therefore $n_7= 50$ (but not 51). More details on these 10 datasets are shown in Table \ref{tab:Table1}. We indicate the number of attributes, the number of rated items with a score of 1, 2, 3, 4, and 5 stars, the average rating, and the standard deviation. 

		\begin{table}[!htb]
			\centering\scriptsize
			\caption{Description of the data sets used in our first experiments}
			\label{tab:Table1}
			\begin{tabular}{ccccccccc}
				\multicolumn{2}{c}{}&\multicolumn{5}{c}{Number of rated items}&average&standard\\
				$u_i$&$n_i$&1$\star$&2$\star$&3$\star$&4$\star$&5$\star$& rating&deviation\\
				\cmidrule(lr){1-1}\cmidrule(lr){2-2}\cmidrule(lr){3-7}\cmidrule(lr){8-8}\cmidrule(lr){9-9}
				$u_1$	& 49	& 5	& 68& 	237& 	169	& 	21& 	3.27& 	0.78\\
				$u_2$	& 50	& 12& 60& 	246& 	177& 	5& 		3.21& 	0.75\\
				$u_3$	& 50	& 7	& 4	& 	278	& 	158	& 	9	& 	3.23& 	0.70\\
				$u_4$	& 48	& 9	& 50& 	141& 	229	& 	71& 	3.61& 	0.91\\
				$u_5$	& 51	& 7	& 43& 	125	& 	166	& 	159	& 	3.85& 1.01\\
				$u_6$	& 48	& 4	& 9& 	41	& 	312	& 	134	& 	4.13& 0.69\\
				$u_7$	& 50	& 17& 40& 	116& 	189& 	138& 	3.78& 	1.04\\
				$u_8$	& 50	& 2	& 15& 	195	& 	214	& 	74	& 	3.69& 0.77\\
				$u_9$	& 49	& 4	& 21& 	115	& 	279	& 	81	& 	3.82& 0.78\\
				$u_{10}$	& 45	& 4	& 47& 	153& 	230	& 	66	& 	3.61& 0.86\\
				\bottomrule
			\end{tabular}
		\end{table}
 
Note that user
 ratings may seem inconsistent. For example, it is not rare that a user gives two
 different ratings to two items having the same attribute values. This can be due to a missing attribute in the system or to human inconsistency as user preferences may be impacted by various contextual aspects or simply changed with time \cite{Said18}. The considered datasets are no exception. For example, three restaurants rated by $u_4$ have the same attribute values while the user has assigned a score of 3, 4 and 5 stars, respectively. Table \ref{tab:Table1} shows that most of the user ratings are of 3 and 4 stars, which is a known phenomenon \cite{Mansoury21}.
 
For each user, we performed a 10-fold cross-validation process with 450 restaurants as a training set $I'$ and the remaining 50 items of $I\setminus I'$ as a test set. The algorithm that minimizes $f_1$ will be called Algo1, while Algo2 is for the minimization of $f_2$. These two algorithms are called Model1 and Model3 in \cite{HKT21}, respectively. As mentioned in Section \ref{sec:3}, the output of Algo1 is a vector $\boldsymbol{x}\in \{0,1\}^{n}$. We have to explain how this output is used to predict the ratings given by the user to the 50 restaurants in the test set.
 
While function $\tau$ translates an $s$-star rating in $\{1,\ldots,s\}$ into a $d$-rating in $[0,n]$, we consider the following inverse function $\tau^{-1}: [0,n]\rightarrow \{1,\ldots,s\}$:
\begin{align*}
\tau^{-1}(\delta)=s-\round*{\frac{\delta(s-1)}{n}},\label{formula2}
\end{align*}
\noindent where $\round{}$ if the nearest integer function. For example, for $n=20$ and $s=5$, a $d$-rating $\delta=7$ is transformed into an $5$-star rating $\tau^{-1}(7)=5-\round*{\frac{28}{20}}=4$. 
 
Consider an item $i\in I\setminus I'$. The Hamming distance $d(\boldsymbol{v}^i,\boldsymbol{x})$ between  the vertex $\boldsymbol{v}^i$ of the hypercube representing the item $i$ and the vertex $\boldsymbol{x}$ representing the user model
is our guess of the number of attribute values in $\boldsymbol{v}^i$ that do not match the user preferences. We therefore transform this distance into an $s$-star rating $\tau^{-1}(d(\boldsymbol{v}^i,\boldsymbol{x}))$ and compare it with the actual rating $r_i$. The average error $F_1(\boldsymbol{x})$ induced by $\boldsymbol{x}$ when predicting the ratings given by the user to the restaurants in $I\setminus I'$ is therefore defined as follows:
 $$F_1(\boldsymbol{x}) =\frac{1}{|I|-|I'|}\sum_{i\in I\setminus I'}|\tau^{-1}(d(\boldsymbol{v}^i,\boldsymbol{x}))-\rho_i|.$$
 
 For Algo2, the average prediction error is calculated in exactly the same way, except that distance $d'$ is used instead of the Hamming distance $d$. The average prediction errors of Algo1 and Algo2 are reported in Table \ref{tab:Table2}. We have run our algorithms on a 3 GHz Intel Xeon X5675 machine with 8 GB of RAM, and all integer programs were solved using CPLEX (v12.2), with a time limit of 1 second. When CPLEX was interrupted before completing the optimization, we report the best solution found. Experiments have shown that no noticeable improvement is obtained with longer computing times, of the order of a minute or an hour.

		\begin{table}[!htb]
			\centering\scriptsize
			\caption{Average prediction errors (in `stars') produced by each method for training sets with 450 items and test sets with 50 items.}
			\label{tab:Table2}
			\begin{tabular}{cccccccc}
				$u_i$	&Algo1&	Algo2&	NB&	RF&	SVR&	DT&	NN\\
				\hline
				$u_1$&	0.554&	0.567&	0.550&	0.563&	0.496&	0.678&	0.648 \\
				$u_2$&	0.542&	0.548&	0.552&	0.541&	0.500&	0.682&	0.605\\
				$u_3$&	0.482&	0.476&	0.418&	0.440&	0.380&	0.588&	0.484\\
				$u_4$&	0.674&	0.678&	0.612&	0.665&	0.604&	0.740&	0.705\\
				$u_5$&	0.778&	0.782&	0.876&	0.757&	0.710&	0.888&	0.851\\
				$u_6$&	0.407&	0.410&	0.406&	0.451&	0.388&	0.612&	0.493\\
				$u_7$&	0.770&	0.768&	0.772&	0.767&	0.716&	0.940&	0.936\\
				$u_8$&	0.614&	0.608&	0.580&	0.533&	0.570&	0.622&	0.653\\
				$u_9$&	0.500&	0.500&	0.498&	0.525&	0.470&	0.616&	0.646 \\
				$u_{10}$&	0.644&	0.615&	0.584&	0.580&	0.508&	0.684&	0.665\\
				\hline Average&		0.5964&	0.5953&	0.5848&	0.5821&	0.5342&	0.7050&	0.6686\\
				\bottomrule
			\end{tabular}
		\end{table}

Since the datasets are very small and concern only a very limited number of users, deep learning techniques cannot be used. Hence, for comparing the above results with other techniques, we have run five machine learning methods  widely used in recommender systems, namely Naïve Bayes (NB), Random Forest (RF), Support Vector Regression (SVR), Decision Tree (DT) and Neural Network (NN), using the Scikit machine learning tool \cite{Scikit}. 
The hyper parameters for these machine learning methods were chosen according to the recommendations given by the scikit-learn website (https://scikit-learn.org/). We also did some experiments to make sure we were getting the best results using the chosen hyper parameters.

As for Algo1 and Algo2, we performed a 10-fold cross validation, using 450 items as a training set and 50 items as a test set. The results are also given in Table \ref{tab:Table2}.

We observe that  SVR  is the best method with the smallest average prediction error of 0.5342, while the DT is the worst with an error of 0.7050. The errors for Algo1 and Algo2 are 0.5964 and 0.5953, respectively, and  NB and RF  have slightly better results. As reported in Table \ref{tab:Table7}, $f$-tests and $t$-tests indicate that the average prediction errors of Algo1, Algo2, NB and RF are not statistically different, while SVR is statistically better than all other methods. This also means that although Algo1 and Algo2 give results that are a bit worse than those produced by NB and RF, there doesn't seem to be any significant advantage to using NB or RF over our two algorithms. It should be noted, however, that the sets $I'$ of rated items are large, which means that the users considered in the above experiments have many interactions with the recommender system. When a user is not very active in the system, the results produced by the various methods can be very different. This is what we want to evaluate now, showing the advantage of Algo1 and Algo2 over other techniques for this cold start situation. 

\begin{table}[!htb]
			\setlength{\extrarowheight}{1pt}
	\centering\scriptsize
	\caption{Results of the statistical tests: $+$ indicates a  significant difference between the methods, while $-$ indicates that the two methods are not significantly different.}
	\label{tab:Table7}
	\begin{tabular}{c|cccccc}
	&	SVR&	NB&		RF&		Algo2&		Algo1&		NN\\
\hline		
		NB&	$+$	&&&&	\\				
		RF&	$+$&	$-$&&&&\\				
		Algo2&	$+$&	$-$&	$-$&&&\\			
		Algo1&	$+$&	$-$&	$-$&	$-$&&\\		
		NN&	$+$&	$+$&		$+$&		$+$&			$+$&	\\
		DT&	$+$&	$+$&		$+$&		$+$&			$+$&			$+$
	\end{tabular}
\end{table}

\begin{figure}[!htb]
		\centering
		\includegraphics[height=5.9cm,width=11.2cm]{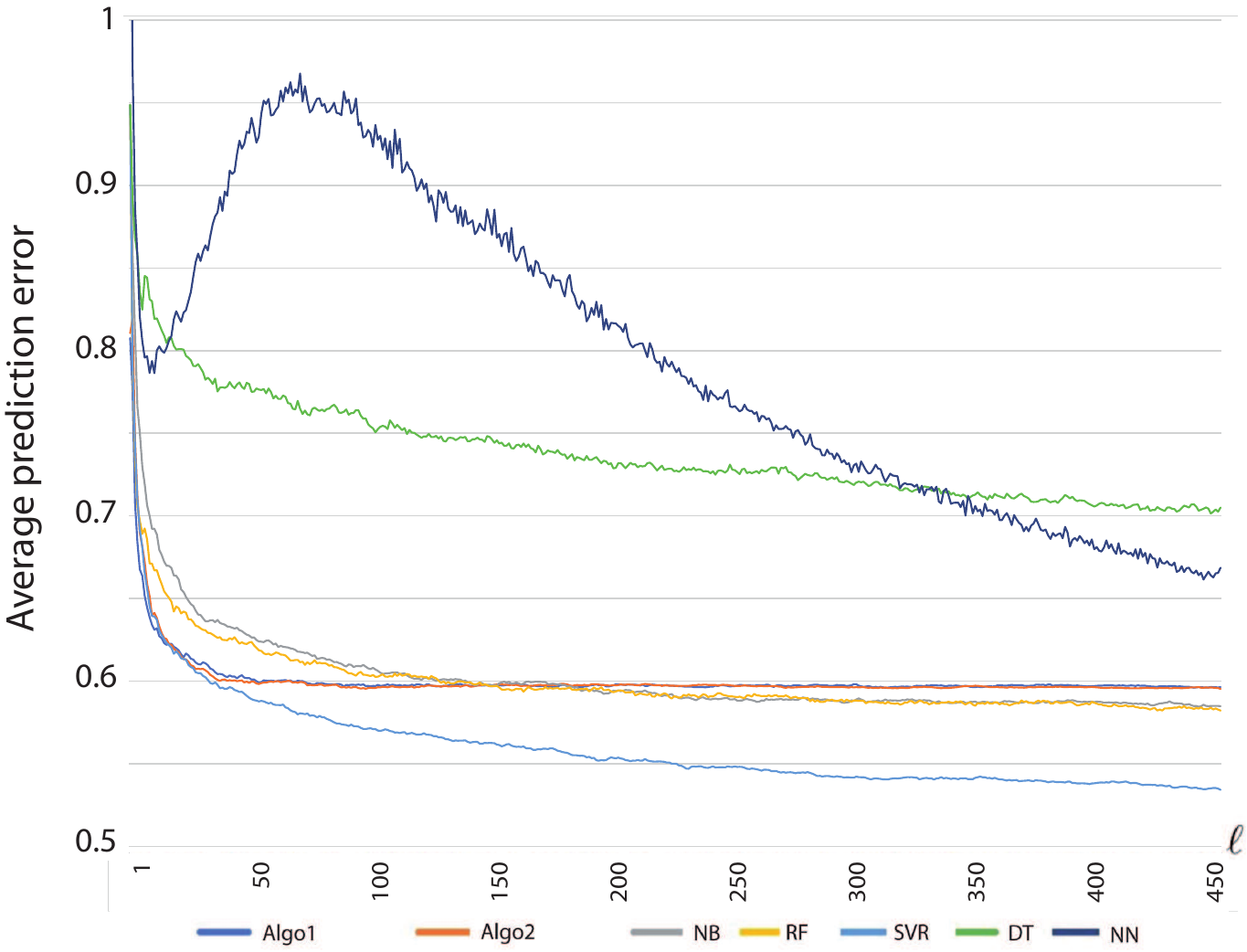}
		\caption{Impact of the size $\ell$ of the training set on the accuracy of each model.}
		\label{fig:Figure2}
	\end{figure}

In the next experiment, we investigate the impact of the size of the training set $I'$ on the accuracy of the recommendations. For each user dataset, we have set aside 10 test sets,  each containing 50 restaurants taken from a different part of the dataset. For each test set, we considered 450 training sets of increasing size, containing $\ell=1,\ldots,450$ items, randomly selected from the  450 restaurants not included in the test set. We again performed a 10-folds cross validation process, applying Algo1, Algo2 and the five other machine learning techniques. The average results over the ten users are shown in Figure \ref{fig:Figure2} and are also reported in Table \ref{tab:Table3}, for $\ell=1$ to 22, and for $\ell=132$ to 150.

Our first observation is that Algo1 and Algo2 produce similar results. Hence, `don't care' values do not seem to have any impact in this experiment.
We therefore only compare the output of Algo1 with those of the five machine learning algorithms. To evaluate whether there is a significant difference between two methods, we used the standard $f$-tests and $t$-tests. When a method performs significantly better or worse than Algo1, we indicate it with a gray cell in Table \ref{tab:Table3}. Smallest average errors are shown with bold characters.
	
		\begin{table}[!htb]
			\setlength{\extrarowheight}{-0.7pt}
			\centering \scriptsize
			\caption{Average prediction error obtained for $\ell=1$ to 22 and for $\ell=132$ to 150. Bold numbers indicate the best results. Gray cells indicate a significant difference with Algo1.}
			\label{tab:Table3}
			\begin{tabular}{cccccccc}
 $\ell$	&Algo1&	Algo2&	NB&	RF&	SVR&	DT&	NN\\
\hline
1	&  	{\bf 0.8076}	&  	0.8105	&  	\cellcolor{gray!60}0.9487	&  	\cellcolor{gray!60}0.9487	&  	\cellcolor{gray!60}0.9487	&  	\cellcolor{gray!60}0.9487	&  	\cellcolor{gray!60}1.1647	\\
2	&  	{\bf 0.7756}	&  	0.8191	&  	\cellcolor{gray!60}0.8591	&  	0.7896	&  	0.7918	&  \cellcolor{gray!60}	0.8913	&  \cellcolor{gray!60}	0.9730	\\
3	&  	{\bf 0.7076}	&  	0.7464	&  	\cellcolor{gray!60}0.8109	&  	0.7356	&  	0.7427	&  \cellcolor{gray!60}0.8684	&  	\cellcolor{gray!60}0.8865	\\
4	&  	{\bf 0.6840}	&  	0.7119	&  	\cellcolor{gray!60}0.7658	&  	0.7053	&  	0.7097	&  \cellcolor{gray!60}	0.8577	&  	\cellcolor{gray!60}0.8571	\\
5	&  	{\bf 0.6676}	&  	0.6897	&  	\cellcolor{gray!60}0.7504	&  	0.6970	&  	0.6916	&  	\cellcolor{gray!60}0.8361	&  	\cellcolor{gray!60}0.8205	\\
6	&  	{\bf 0.6637}	&  	0.6815	&  \cellcolor{gray!60}	0.7297	&  	0.6891	&  	0.6810	&  	\cellcolor{gray!60}0.8247	&  	\cellcolor{gray!60}0.8057	\\
7	&  	{\bf 0.6515}	&  	0.6720	&  	\cellcolor{gray!60}0.7192	&  	\cellcolor{gray!60}0.6924	&  	0.6652	&  	\cellcolor{gray!60}0.8452	&  	\cellcolor{gray!60}0.7957	\\
8	&  {\bf 0.6446}	&  	0.6589	&  	\cellcolor{gray!60}0.7063	&  	\cellcolor{gray!60}0.6856	&  	0.6559	&  	\cellcolor{gray!60}0.8442	&  \cellcolor{gray!60}	0.7966	\\
9	&  	{\bf 0.6387}	&  	0.6514	&  	\cellcolor{gray!60}0.7001	&  	0.6712	&  	0.6453	&  	\cellcolor{gray!60}0.8312	&  	\cellcolor{gray!60}0.7865	\\
10	&  	{\bf 0.6341}	&  	0.6397	&  	\cellcolor{gray!60}0.6922	&  	0.6711	&  	0.6394	&  	\cellcolor{gray!60}0.8298	&  	\cellcolor{gray!60}0.7937	\\
11	&  {\bf 	0.6311}	&  	0.6413	&  	\cellcolor{gray!60}0.6922	&  	0.6670	&  	0.6391	&  	\cellcolor{gray!60}0.8191	&  	\cellcolor{gray!60}0.7863	\\
12	&  {\bf 	0.6319}	&  	0.6381	&  	\cellcolor{gray!60}0.6888	&  	0.6671	&  	0.6371	&  	\cellcolor{gray!60}0.8195	&  	\cellcolor{gray!60}0.8002	\\
13	&  	{\bf 0.6271}	&  	0.6331	&  	\cellcolor{gray!60}0.6794	&  	0.6625	&  	0.6305	&  	\cellcolor{gray!60}0.8160	&  	\cellcolor{gray!60}0.8026	\\
14	&  	{\bf 0.6260}	&  	0.6291	&  	\cellcolor{gray!60}0.6764	&  	0.6587	&  	0.6279	&  	\cellcolor{gray!60}0.8123	&  	\cellcolor{gray!60}0.8000	\\
15	&  	{\bf 0.6230}	&  	0.6261	&  	\cellcolor{gray!60}0.6723	&  	0.6544	&  	0.6253	&  	\cellcolor{gray!60}0.8090	&  	\cellcolor{gray!60}0.7987	\\
16	&  {\bf 	0.6220}	&  	0.6254	&  	\cellcolor{gray!60}0.6698	&  	0.6523	&  	0.6231	&  	\cellcolor{gray!60}0.8047	&  \cellcolor{gray!60}	0.8016	\\
17	&  	0.6225	&  	0.6224	&  	\cellcolor{gray!60}0.6697	&  	0.6504	&  {\bf 	0.6217}	&  	\cellcolor{gray!60}0.8081	&  \cellcolor{gray!60}	0.8059	\\
18	&  	0.6226	&  	0.6219	&  	\cellcolor{gray!60}0.6685	&  	0.6480	&  {\bf 	0.6202}	&  	\cellcolor{gray!60}0.8075	&  	\cellcolor{gray!60}0.8089	\\
19	&  	0.6207	&  	0.6196	&  	\cellcolor{gray!60}0.6639	&  	0.6421	&  	{\bf 0.6165}	&  	\cellcolor{gray!60}0.8029	&  	\cellcolor{gray!60}0.8190	\\
20	&  	0.6199	&  	0.6182	&  	\cellcolor{gray!60}0.6637	&  	0.6452	&  	{\bf 0.6180}	&  	\cellcolor{gray!60}0.8006	&  	\cellcolor{gray!60}0.8238	\\
21	&  	0.6185	&  	0.6174	&  	\cellcolor{gray!60}0.6615	&  	0.6441	&  	{\bf 0.6163}	&  \cellcolor{gray!60}	0.8009	&  	\cellcolor{gray!60}0.8204	\\
22	&  	0.6162	&  	0.6146	&  	0.6554	&  	0.6409	&  {\bf 	0.6138}	&  \cellcolor{gray!60}	0.8009	&  	\cellcolor{gray!60}0.8174	\\
\multicolumn{8}{c}{$\vdots$}  		\\
132	&  	0.5973	&  	0.5968	&  	0.6007	&  	0.5994	&  	{\bf 0.5648}	&  	\cellcolor{gray!60}0.7471	&  	\cellcolor{gray!60}0.8861	\\
133	&  	0.5980	&  	0.5967	&  	0.6011	&  	0.5985	&  	\cellcolor{gray!60}{\bf 0.5642}	&  	\cellcolor{gray!60}0.7472	&  	\cellcolor{gray!60}0.8841	\\
134	&  	0.5981	&  	0.5969	&  	0.6009	&  	0.5984	&  	\cellcolor{gray!60}{\bf 0.5637}	&  \cellcolor{gray!60}	0.7470	&  	\cellcolor{gray!60}0.8841	\\
135	&  	0.5982	&  	0.5970	&  	0.6012	&  	0.5990	&  	\cellcolor{gray!60}{\bf 0.5641}	&  	\cellcolor{gray!60}0.7458	&  	\cellcolor{gray!60}0.8879	\\
136	&  	0.5980	&  	0.5971	&  	0.6010	&  	0.5998	&  \cellcolor{gray!60}{\bf 0.5639}	&  	\cellcolor{gray!60}0.7466	&  	\cellcolor{gray!60}0.8771	\\
137	&  	0.5983	&  	0.5966	&  	0.6009	&  	0.5999	&  	\cellcolor{gray!60}{\bf 0.5640}	&  	\cellcolor{gray!60}0.7476	&  	\cellcolor{gray!60}0.8869	\\
138	&  	0.5977	&  	0.5971	&  	0.5999	&  	0.5996	&  \cellcolor{gray!60}{\bf 0.5640}	&  	\cellcolor{gray!60}0.7473	&  	\cellcolor{gray!60}0.8752	\\
139	&  	0.5978	&  	0.5974	&  	0.6003	&  	0.5992	&  \cellcolor{gray!60}{\bf 0.5642}	&  	\cellcolor{gray!60}0.7477	&  	\cellcolor{gray!60}0.8849	\\
140	&  	0.5974	&  	0.5974	&  	0.5994	&  	0.5991	&  \cellcolor{gray!60}{\bf 0.5632}	&  	\cellcolor{gray!60}0.7451	&  	\cellcolor{gray!60}0.8764	\\
141	&  	0.5971	&  	0.5966	&  	0.5987	&  	0.5993	&  	\cellcolor{gray!60}{\bf 0.5628}	&  	\cellcolor{gray!60}0.7453	&  	\cellcolor{gray!60}0.8776	\\
142	&  	0.5971	&  	0.5967	&  	0.5982	&  	0.5983	&  	\cellcolor{gray!60}{\bf 0.5631}	&  	\cellcolor{gray!60}0.7464	&  	\cellcolor{gray!60}0.8788	\\
143	&  	0.5973	&  	0.5969	&  	0.5973	&  	0.5979	&  	\cellcolor{gray!60}{\bf 0.5632}	&  	\cellcolor{gray!60}0.7464	&  	\cellcolor{gray!60}0.8704	\\
144	&  	0.5974	&  	0.5973	&  	0.5978	&  	0.5983	&  \cellcolor{gray!60}{\bf 0.5632}	&  	\cellcolor{gray!60}0.7473	&  	\cellcolor{gray!60}0.8726	\\
145	&  	0.5974	&  	0.5974	&  	0.5991	&  	0.5991	&  	\cellcolor{gray!60}{\bf 0.5621}	&  	\cellcolor{gray!60}0.7462	&  	\cellcolor{gray!60}0.8769	\\
146	&  	0.5972	&  	0.5976	&  	0.5987	&  	0.5976	&  	\cellcolor{gray!60}{\bf 0.5621}	&  	\cellcolor{gray!60}0.7460	&  	\cellcolor{gray!60}0.8740	\\
147	&  	0.5974	&  	0.5977	&  	0.5977	&  	0.5980	&  	\cellcolor{gray!60}{\bf 0.5626}	&  	\cellcolor{gray!60}0.7447	&  	\cellcolor{gray!60}0.8727	\\
148	&  	0.5976	&  	0.5974	&  	0.5977	&  	0.5976	&  	\cellcolor{gray!60}{\bf 0.5625}	&  	\cellcolor{gray!60}0.7482	&  	\cellcolor{gray!60}0.8856	\\
149	&  	0.5970	&  	0.5972	&  	0.5978	&  	0.5969	&  	\cellcolor{gray!60}{\bf 0.5626}	&  	\cellcolor{gray!60}0.7478	&  	\cellcolor{gray!60}0.8787	\\
150	&  	0.5969	&  	0.5974	&  	0.5975	&  	0.5968	&  	\cellcolor{gray!60}{\bf 0.5625}	&  	\cellcolor{gray!60}0.7462	&  	\cellcolor{gray!60}0.8701	\\
\bottomrule 
\end{tabular}
\end{table}

The results shown in Table \ref{tab:Table3} can be divided into six main parts. For $\ell=1$, Algo1 and Algo2 are significantly better than all other methods. For $\ell\in\{2,\ldots,16\}$, the average error of Algo1 is the smallest and significantly smaller than that of NB, DT and NN. However, even if the results look better, they are not significantly different from the results of RF (with 2 exceptions) or SVR. For $\ell\in\{17,\ldots,450\}$, SVR got the smallest average error. Nevertheless, for rating sets containing less than 133 ratings, the error produced by SVR is not significantly smaller than that of Algo1 which produced the second best result. We also observe that the prediction errors produced with DT and NN are significantly bigger than those of Algo1 for all values $\ell=1,\ldots,450$. For $\ell> 148$, RF becomes the method with the second best result, slightly better than that of Algo1 and Algo2. This is summarized in Table \ref{tab:Table4}.

		\begin{table}[!htb]
			\centering \scriptsize
			\caption{Categories of results according to the size $\ell$ of the training set.}
			\label{tab:Table4}
			\begin{tabular}{cccc}
				$\ell$&	Similar error as Algo1&	Smallest error &	Second best error\\
				\hline
				$=1$&	Algo2&	Algo1&	Algo2\\
				\hline\multirow{3}{*}{$\in\{2,\ldots,16\}$}&	Algo2 &	&\\
				&	SVR &Algo1	& SVR (1
				exception)	\\
				&	RF (2 exceptions)	&&\\
				\hline\multirow{3}{*}{$\in\{17,\ldots,21\}$}&	Algo2&	&\\
				&	SVR & SVR & Algo2 \\
				&RF&&\\
				\hline\multirow{4}{*}{$\in\{22,\ldots,132\}$}& Algo2&	\multirow{4}{*}{SVR}&\multirow{4}{*}{Algo2 (4 exceptions)}\\
				&	SVR &  &  \\
				&RF&&\\
				&NB&&\\
				\hline\multirow{3}{*}{$\in\{133,\ldots,148\}$}&	Algo2&	&\\
				&	RF & SVR & Algo2 \\
				&NB&&\\
				\hline\multirow{3}{*}{$\in\{149,\ldots,450\}$}&	Algo2&	&\\
				&	RF & SVR & RF \\
				&NB&&\\
				\bottomrule 
			\end{tabular}
		\end{table}

\begin{figure}[!htb]
		\centering
		\includegraphics[width=0.60\textwidth,keepaspectratio]{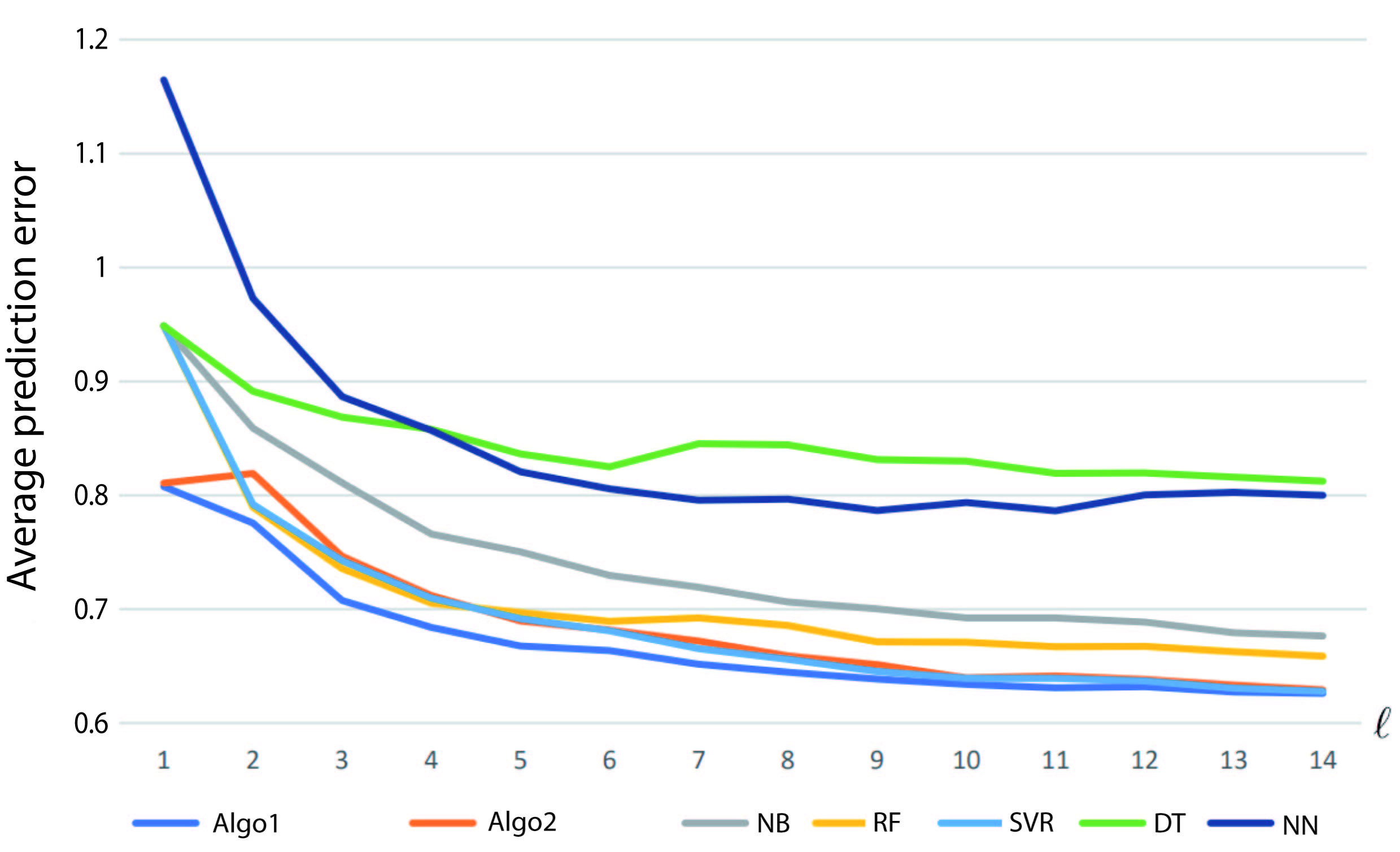}
		\caption{Impact of the size of the training set on the accuracy of each model.}
		\label{fig:Figure3}
	\end{figure}

In summary, with respect to the cold start problem, Algo1 and Algo2 produce relatively small prediction errors for small values of $\ell$. As illustrated in Figure \ref{fig:Figure2}, the prediction error of Algo1 and Algo2 reaches its lowest value around $\ell=50$, without a significant improvement for larger training sets. The SVR method, on the other hand, reaches the lowest error value for larger training sets, with improved prediction accuracy as $\ell$ increases. A zoom on the same curves for $\ell\leq 14$ is shown in Figure \ref{fig:Figure3}.\\

\subsection{Experiments with movie datasets}

As next experiment, we consider movie datasets taken from the MovieLens website\linebreak[10] (https://movielens.org) which allows to characterize each movie with a large set of attributes. We have extracted ten datasets, each containing 50 movies that were rated by a single user on a 5-star scale, with half-star increments (0.5 stars - 5.0 stars). Each movie is characterized using 1,128 attributes which exhibit particular properties like movie's genre (action, crime, drama etc.), movie's theme (political corruption, midlife crisis, racism, memory loss etc.), awards given to the movie (Oscar awards in different categories, Saturn award for best special effects etc.), famous directors (Spielberg, Tarantino etc.), and other characteristics  (true story, allegory, twist ending, thought provoking etc.) as well as viewers personal impressions (too long, unfunny, scary etc.). 

As for restaurants, the dataset corresponding to user $u_i$, is characterized by $n_i$ attributes from the full attributes list where, an attribute is included in the dataset only if it exists in at least one of the movies rated by the user. As a result, the numbers $n_i$ of attributes vary from 627 to 796.  
More details on the 10 datasets are shown in Table \ref{tab:Table8} where we indicate, for each user, the number of attributes, the number of rated items with each score, the average rating, and the standard deviation. We notice that most ratings distribution are between 3.0 and 4.0. 

		\begin{table}[!htb]
			\centering\scriptsize
			\caption{Description of the datasets used in our experiments with movies}
			\label{tab:Table8}
\setlength{\tabcolsep}{4pt}
			\begin{tabular}{cccccccccccccc}
				\multicolumn{2}{c}{}&\multicolumn{10}{c}{Number of rated items}&average&standard\\
				$u_i$&$n_i$&0.5$\star$&1.0$\star$&1.5$\star$&2.0$\star$&2.5$\star$&3.0$\star$&3.5$\star$&4.0$\star$&4.5$\star$&5.0$\star$& rating&deviation\\
				\cmidrule(lr){1-1}\cmidrule(lr){2-2}\cmidrule(lr){3-12}\cmidrule(lr){13-13}\cmidrule(lr){14-14}
				$u_1$	& 753&	1&	0&	0&	1&	5&	13&	10&	15&	3&	2&	3.45&	0.79\\
				$u_2$&	696&	0&	4&	0&	9&	0&	20&	0&	12&	0&	5&	3.10&	1.07\\
				$u_3$&	735&	0&	0&	1&	4&	6&	11&	15&	11&	2&	0&	3.26&	0.69\\
				$u_4$&	748&	0&	2&	0&	3&	0&	25&	0&	12&	0&	8&	3.42&	0.97\\
				$u_5$&	627&	0&	0&	0&	1&	2&	25&	6&	12&	0&	4&	3.42&	0.67\\
				$u_6$&	649&	0&	1&	0&	2&	0&	17&	0&	26&	0&	4&	3.60&	0.78\\
				$u_7$&	745&	0&	0&	0&	1&	0&	18&	0&	26&	0&	5&	3.70&	0.68\\
				$u_8$&	712&	0&	6&	0&	5&	0&	30&	0&	9&	0&	0&	2.84&	0.87\\
				$u_9$&	796&	6&	4&	0&	6&	5&	5&	7&	6&	4&	7&	2.96&	1.47\\
				$u_{10}$&	755&	1&	1&	3&	2&	6&	11&	10&	8&	4&	4&	3.26&	1.04\\
				\bottomrule
			\end{tabular}
		\end{table}

For each user dataset, we have set aside 10 test sets,  each containing 5 movies taken from a different part of the dataset. For each test set, we considered 45 training sets of increasing size, containing $\ell=1,\ldots,45$ items, randomly selected from the  45 movies not included in the test set. We again performed a 10-folds cross validation process, applying Algo1, Algo2 and the five other machine learning techniques. As shown in \cite{HKT21}, the numbers of constraints and variables in the integer programs of Algo1 and Algo2 increase linearly with the number of attributes, and we have therefore set the time limit for CPLEX at two seconds, to ensure that at least one feasible solution can be generated. Here again, no noticeable improvement can be obtained with longer computing times.  

	\begin{figure}[!htb]
		\centering
		\includegraphics[height=9cm,width=13cm]{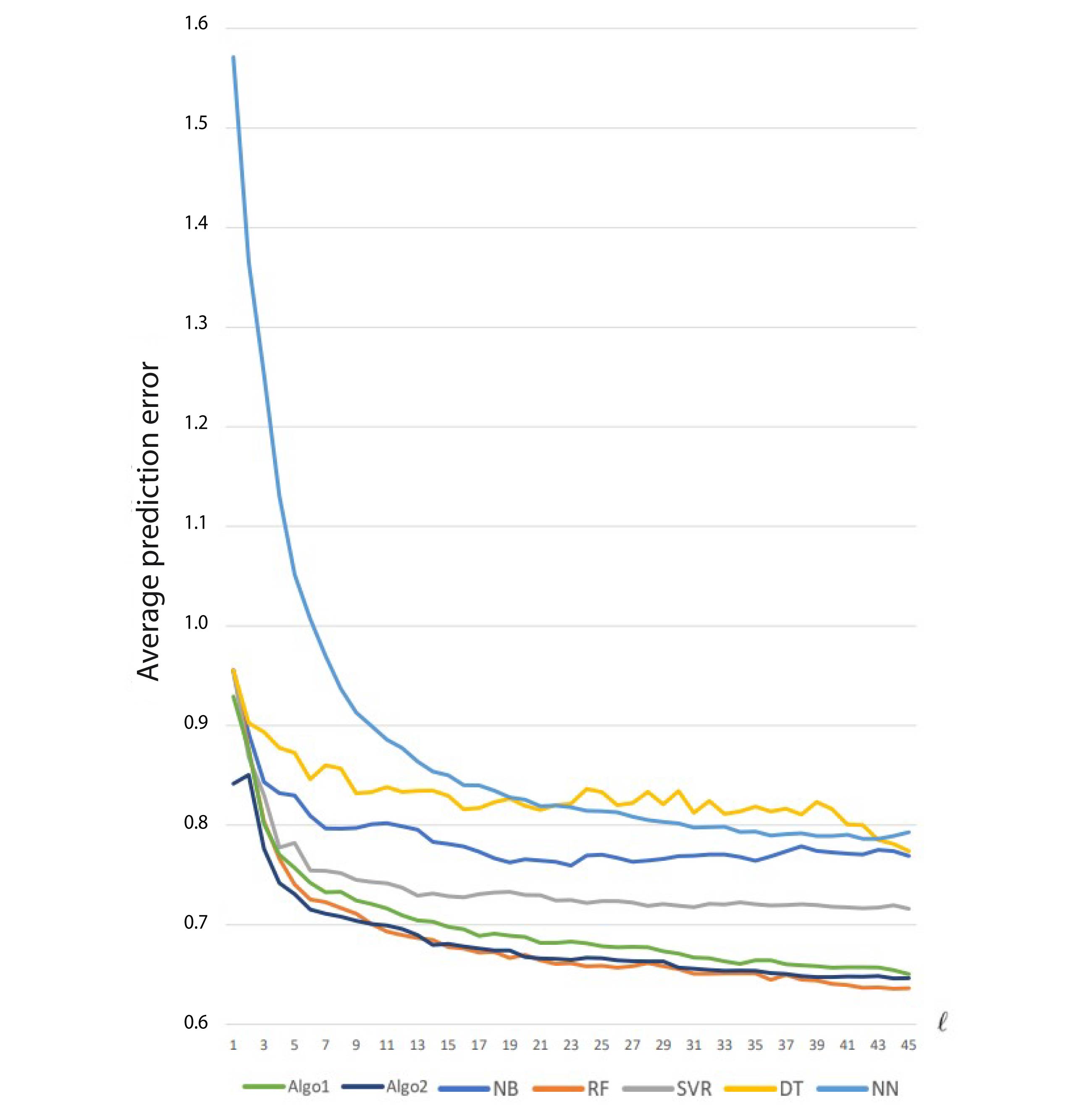}
		\caption{Impact of the size $\ell$ of the training set on the accuracy of each model for 10 movie datasets.}
		\label{fig:Figure5}
	\end{figure}
	
\begin{table}[!htb]
			\setlength{\extrarowheight}{-0.3pt}
\centering \scriptsize
\caption{Average prediction error obtained for $\ell=1$ to 45 for 10 movie datasets. Bold numbers indicate the best results. Gray cells indicate a significant difference with Algo2.}
\label{tab:Table9}
\begin{tabular}{cccccccc}
$\ell$	&Algo1&	Algo2&	NB&	RF&	SVR&	DT&	NN\\
\hline
1&	\cellcolor{gray!60}0.9292&	{\bf 0.8415}&	\cellcolor{gray!60}0.9555&	\cellcolor{gray!60}0.9555&	\cellcolor{gray!60}0.9555&	\cellcolor{gray!60}0.9555&	\cellcolor{gray!60}1.5712\\
2&	0.8754&	{\bf 0.8505}&	\cellcolor{gray!60}0.8919&	0.8757&	0.8682&	\cellcolor{gray!60}0.9026&	\cellcolor{gray!60}1.3658\\
3&	0.8012&	{\bf 0.7762}&	\cellcolor{gray!60}0.8433&	\cellcolor{gray!60}0.8037&	\cellcolor{gray!60}0.8303&	\cellcolor{gray!60}0.8934&	\cellcolor{gray!60}1.2524\\
4&	\cellcolor{gray!60}0.7704&	{\bf 0.7419}&	\cellcolor{gray!60}0.8322&	0.7663&	\cellcolor{gray!60}0.7776&	\cellcolor{gray!60}0.8777&	\cellcolor{gray!60}1.1306\\
5&	\cellcolor{gray!60}0.7570&	{\bf 0.7307}&	\cellcolor{gray!60}0.8298&	0.7404&	\cellcolor{gray!60}0.7822&	\cellcolor{gray!60}0.8727&	\cellcolor{gray!60}1.0515\\
6&	\cellcolor{gray!60}0.7422&	{\bf 0.7153}&	\cellcolor{gray!60}0.8093&	0.7255&	\cellcolor{gray!60}0.7546&	\cellcolor{gray!60}0.8461&	\cellcolor{gray!60}1.0072\\
7&	0.7327&	{\bf 0.7111}&	\cellcolor{gray!60}0.7968&	0.7228&	\cellcolor{gray!60}0.7541&	\cellcolor{gray!60}0.8602&	\cellcolor{gray!60}0.9699\\
8&	0.7331&	{\bf 0.7082}&	\cellcolor{gray!60}0.7965&	0.7169&	\cellcolor{gray!60}0.7518&	\cellcolor{gray!60}0.8568&	\cellcolor{gray!60}0.9370\\
9&	0.7244&	{\bf 0.7040}&	\cellcolor{gray!60}0.7972&	0.7111&	\cellcolor{gray!60}0.7450&	\cellcolor{gray!60}0.8319&	\cellcolor{gray!60}0.9131\\
10&	0.7208&	{\bf 0.7007}&	\cellcolor{gray!60}0.8010&	0.7008&	\cellcolor{gray!60}0.7430&	\cellcolor{gray!60}0.8331&	\cellcolor{gray!60}0.8996\\
11&	0.7164&	0.6993&	\cellcolor{gray!60}0.8019&	{\bf 0.6931}&	\cellcolor{gray!60}0.7416&	\cellcolor{gray!60}0.8382&	\cellcolor{gray!60}0.8859\\
12&	0.7095&	0.6958&	\cellcolor{gray!60}0.7988&	{\bf 0.6895}&	\cellcolor{gray!60}0.7373&	\cellcolor{gray!60}0.8334&	\cellcolor{gray!60}0.8775\\
13&	0.7044&	0.6895&	\cellcolor{gray!60}0.7954&	{\bf 0.6867}&	\cellcolor{gray!60}0.7292&	\cellcolor{gray!60}0.8344&	\cellcolor{gray!60}0.8639\\
14&	0.7031&	{\bf 0.6797}&	\cellcolor{gray!60}0.7832&	0.6848&	\cellcolor{gray!60}0.7313&	\cellcolor{gray!60}0.8347&	\cellcolor{gray!60}0.8538\\
15&	0.6979&	0.6807&	\cellcolor{gray!60}0.7812&	{\bf 0.6776}&	\cellcolor{gray!60}0.7287&	\cellcolor{gray!60}0.8293&	\cellcolor{gray!60}0.8500\\
16&	0.6954&	0.6783&	\cellcolor{gray!60}0.7786&	{\bf 0.6762}&	\cellcolor{gray!60}0.7276&	\cellcolor{gray!60}0.8159&	\cellcolor{gray!60}0.8402\\
17&	0.6887&	0.6762&	\cellcolor{gray!60}0.7732&	{\bf 0.6722}&	\cellcolor{gray!60}0.7308&	\cellcolor{gray!60}0.8172&	\cellcolor{gray!60}0.8400\\
18&	0.6911&	0.6742&	\cellcolor{gray!60}0.7667&	{\bf 0.6727}&	\cellcolor{gray!60}0.7323&	\cellcolor{gray!60}0.8231&	\cellcolor{gray!60}0.8346\\
19&	0.6890&	0.6742&	\cellcolor{gray!60}0.7625&	{\bf 0.6667}&	\cellcolor{gray!60}0.7330&	\cellcolor{gray!60}0.8267&	\cellcolor{gray!60}0.8280\\
20&	0.6876&	{\bf 0.6677}&	\cellcolor{gray!60}0.7657&	0.6697&	\cellcolor{gray!60}0.7299&	\cellcolor{gray!60}0.8194&	\cellcolor{gray!60}0.8256\\
21&	0.6817&	0.6661&	\cellcolor{gray!60}0.7645&	{\bf 0.6644}&	\cellcolor{gray!60}0.7296&	\cellcolor{gray!60}0.8152&	\cellcolor{gray!60}0.8190\\
22&	0.6819&	0.6657&	\cellcolor{gray!60}0.7631&	{\bf 0.6608}&	\cellcolor{gray!60}0.7243&	\cellcolor{gray!60}0.8202&	\cellcolor{gray!60}0.8197\\
23&	0.6830&	0.6647&	\cellcolor{gray!60}0.7594&	{\bf 0.6613}&	\cellcolor{gray!60}0.7249&	\cellcolor{gray!60}0.8216&	\cellcolor{gray!60}0.8180\\
24&	0.6814&	0.6669&	\cellcolor{gray!60}0.7694&	{\bf 0.6581}&	\cellcolor{gray!60}0.7218&	\cellcolor{gray!60}0.8363&	\cellcolor{gray!60}0.8144\\
25&	0.6785&	0.6665&	\cellcolor{gray!60}0.7704&	{\bf 0.6588}&	\cellcolor{gray!60}0.7237&	\cellcolor{gray!60}0.8332&	\cellcolor{gray!60}0.8140\\
26&	0.6774&	0.6642&	\cellcolor{gray!60}0.7672&	{\bf 0.6569}&	\cellcolor{gray!60}0.7237&	\cellcolor{gray!60}0.8200&	\cellcolor{gray!60}0.8130\\
27&	0.6777&	0.6634&	\cellcolor{gray!60}0.7632&	{\bf 0.6584}&	\cellcolor{gray!60}0.7222&	\cellcolor{gray!60}0.8223&	\cellcolor{gray!60}0.8085\\
28&	0.6773&	0.6633&	\cellcolor{gray!60}0.7644&	{\bf 0.6616}&	\cellcolor{gray!60}0.7190&	\cellcolor{gray!60}0.8333&	\cellcolor{gray!60}0.8051\\
29&	0.6734&	0.6632&	\cellcolor{gray!60}0.7661&	{\bf 0.6584}&	\cellcolor{gray!60}0.7209&	\cellcolor{gray!60}0.8209&	\cellcolor{gray!60}0.8033\\
30&	0.6710&	0.6569&	\cellcolor{gray!60}0.7689&	{\bf 0.6554}&	\cellcolor{gray!60}0.7191&	\cellcolor{gray!60}0.8342&	\cellcolor{gray!60}0.8017\\
31&	0.6670&	0.6558&	\cellcolor{gray!60}0.7693&	{\bf 0.6509}&	\cellcolor{gray!60}0.7177&	\cellcolor{gray!60}0.8123&	\cellcolor{gray!60}0.7976\\
32&	0.6665&	0.6547&	\cellcolor{gray!60}0.7705&	{\bf 0.6508}&	\cellcolor{gray!60}0.7210&	\cellcolor{gray!60}0.8244&	\cellcolor{gray!60}0.7980\\
33&	0.6632&	0.6538&	\cellcolor{gray!60}0.7706&	{\bf 0.6513}&	\cellcolor{gray!60}0.7203&	\cellcolor{gray!60}0.8113&	\cellcolor{gray!60}0.7983\\
34&	0.6608&	0.6540&	\cellcolor{gray!60}0.7681&	{\bf 0.6512}&	\cellcolor{gray!60}0.7224&	\cellcolor{gray!60}0.8137&	\cellcolor{gray!60}0.7932\\
35&	0.6643&	0.6537&	\cellcolor{gray!60}0.7642&	{\bf 0.6512}&	\cellcolor{gray!60}0.7205&	\cellcolor{gray!60}0.8187&	\cellcolor{gray!60}0.7936\\
36&	0.6643&	0.6514&	\cellcolor{gray!60}0.7686&	{\bf 0.6447}&	\cellcolor{gray!60}0.7194&	\cellcolor{gray!60}0.8137&	\cellcolor{gray!60}0.7896\\
37&	0.6604&	0.6506&	\cellcolor{gray!60}0.7737&	{\bf 0.6497}&	\cellcolor{gray!60}0.7197&	\cellcolor{gray!60}0.8166&	\cellcolor{gray!60}0.7910\\
38&	0.6593&	0.6484&	\cellcolor{gray!60}0.7788&	{\bf 0.6449}&	\cellcolor{gray!60}0.7206&	\cellcolor{gray!60}0.8107&	\cellcolor{gray!60}0.7918\\
39&	0.6584&	0.6475&	\cellcolor{gray!60}0.7742&	{\bf 0.6441}&	\cellcolor{gray!60}0.7197&	\cellcolor{gray!60}0.8232&	\cellcolor{gray!60}0.7891\\
40&	0.6570&	0.6474&	\cellcolor{gray!60}0.7726&	{\bf 0.6407}&	\cellcolor{gray!60}0.7181&	\cellcolor{gray!60}0.8161&	\cellcolor{gray!60}0.7892\\
41&	0.6573&	0.6481&	\cellcolor{gray!60}0.7715&	{\bf 0.6394}&	\cellcolor{gray!60}0.7174&	\cellcolor{gray!60}0.8006&	\cellcolor{gray!60}0.7903\\
42&	0.6574&	0.6475&	\cellcolor{gray!60}0.7705&	{\bf 0.6366}&	\cellcolor{gray!60}0.7165&	\cellcolor{gray!60}0.8001&	\cellcolor{gray!60}0.7863\\
43&	0.6572&	0.6483&	\cellcolor{gray!60}0.7751&	{\bf 0.6371}&	\cellcolor{gray!60}0.7173&	\cellcolor{gray!60}0.7850&	\cellcolor{gray!60}0.7863\\
44&	0.6544&	0.6426&	\cellcolor{gray!60}0.7739&	{\bf 0.6358}&	\cellcolor{gray!60}0.7196&	\cellcolor{gray!60}0.7811&	\cellcolor{gray!60}0.7891\\
45&	0.6504&	0.6422&	\cellcolor{gray!60}0.7690&	{\bf 0.6363}&	\cellcolor{gray!60}0.7160&	\cellcolor{gray!60}0.7740&	\cellcolor{gray!60}0.7930\\
\bottomrule							 
\end{tabular}
\end{table}

The average results over the ten users are shown in Figure \ref{fig:Figure5} and are also reported in Table \ref{tab:Table9}. The results of the statistical tests for $\ell \in \{1,\ldots,45\}$, which were performed using the standard $f$-tests and $t$-tests, are represented in Table \ref{tab:Table9} as follows: when a method performs significantly better or worse than Algo2, we indicate it with a gray cell. Best average errors are shown with bold characters.
Note that the results generated by each of the methods are compared to those of Algo2 (instead of Algo1) which got better results, though not much different than Algo1, and also the smallest average error of all the methods for $\ell \in \{1,\ldots,10\}$. A summary of the results appears in Table \ref{tab:Table10}.

		\begin{table}[!htb]
			\centering \scriptsize
			\caption{Categories of results according to the size $\ell$ of the training set.}
			\label{tab:Table10}
			\begin{tabular}{cccc}
				$\ell$&	Similar error as Algo2&	Smallest error &	Second best error\\
				\hline 1&-&Algo2&Algo1\\
				\hline\multirow{3}{*}{2}&	Algo1 &	\multirow{3}{*}{Algo2}&\multirow{3}{*}{SVR}\\
				&	RF 	& & \\
				&	SVR 	& & \\
				\hline 3&Algo1&Algo2&Algo1\\
				\hline$\in\{4,5,6\}$&	RF&	Algo2&RF\\
				\hline\multirow{2}{*}{$\in\{7,8,9,10,14,20\}$}& Algo1&	\multirow{2}{*}{Algo2}&\multirow{2}{*}{RF}\\
				&	RF &  &  \\
				\hline\multirow{2}{*}{$\in\{11,\ldots,45\}\setminus\{14,20\}$}&	Algo1&
				\multirow{2}{*}{RF}&\multirow{2}{*}{Algo2}\\
				&	RF &  &  \\
				\bottomrule 
			\end{tabular}
		\end{table}
	
We can observe that the performance of Algo2 is significantly better than that of NB, DT and NN for $\ell\geq 1$, and of SVR for $\ell\neq 2$. Also, Algo1 is  significantly better than NB, SVR, DT and NN for $\ell\geq 7$.
Nevertheless, for  $\ell \geq 11$ (with 2 exceptions),  the RF method got the best results which are however not significantly different from those of Algo1 and Algo2. 


To test Algo1 and Algo2 on a larger number of instances, we have extracted 1,000 datasets from the MovieLens website (https://movielens.org), each containing 50 movies that were rated by a single user on a 5-star scale, with half-star increments (0.5 stars - 5.0 stars). As a result, the numbers $n_i$ of attributes vary from 440 to 844. 
Note that collaborative filtering methods are usually evaluated using more users than content based methods which in turn, need a lot of data to describe item attributes.
As above, for each user dataset, we have set aside 10 test sets,  each containing 5 movies taken from a different part of the dataset. For each test set, we considered 45 training sets of increasing size, containing $\ell=1,\ldots,20$ items, randomly selected from the  45 movies not included in the test set. Larger sizes $\ell$ for the training set are less relevant to the cold start problem. We again performed a 10-folds cross validation process, applying Algo1, Algo2 and the two best standard machine learning techniques for the cold start problem, namely RF and SVR.

\begin{figure}[!htb]
		\centering
		\includegraphics[height=6.5cm, width=8.5cm]{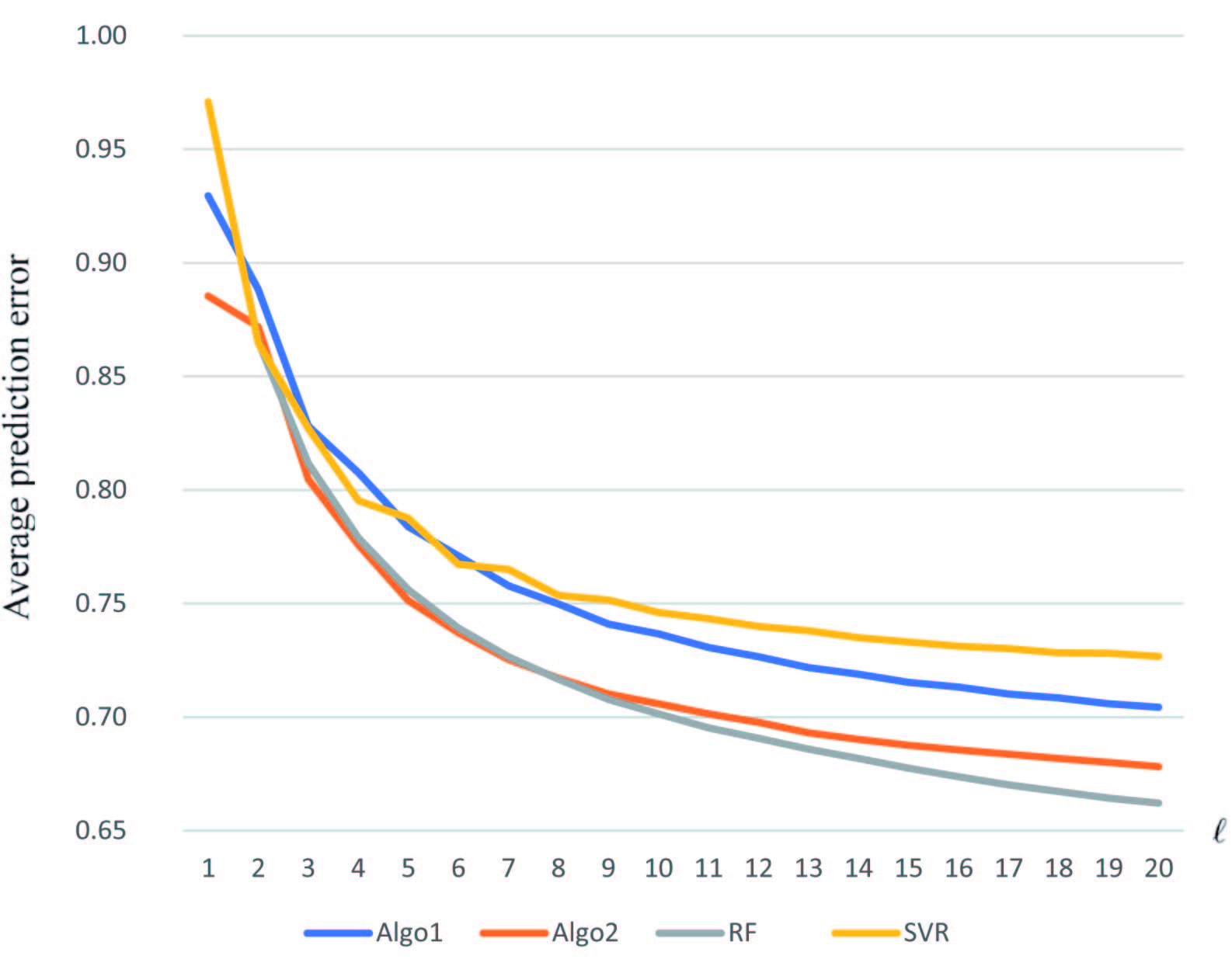}
		\caption{Impact of the size $\ell$ of the training set on the accuracy of each model for 1,000 movie datasets.}
		\label{fig:Figure6}
	\end{figure}

The results are shown in Figure \ref{fig:Figure6} and are also reported in Table \ref{tab:Table11}. When a method performs significantly better or worse than Algo2, we indicate it with a gray cell. Best average errors are shown with bold characters.

\begin{table}[!htb]
						\setlength{\extrarowheight}{-1pt}
	\centering \scriptsize
	\caption{Average prediction error obtained for $\ell=1$ to 20 for 1,000 movie datasets. Bold numbers indicate the best results. Gray cells indicate a significant difference with Algo2.}
	\label{tab:Table11}
	\begin{tabular}{ccccc}
		$\ell$	&Algo1&	Algo2&	RF&	SVR\\
		\hline
		1	&	\cellcolor{gray!60}0.9294	&	{\bf 0.8854}	&	\cellcolor{gray!60}0.9708	&	\cellcolor{gray!60}0.9708	\\
		2	&	0.888251	&	0.8719	&	{\bf 0.8653}	&	0.8646	\\
		3	&	0.828108	&	{\bf 0.8047}	&	0.8116	&	0.827	\\
		4	&	\cellcolor{gray!60}0.8076	&	{\bf 0.7758}	&	0.7788	&	0.7953	\\
		5	&	\cellcolor{gray!60}0.7839	&	{\bf 0.7514}	&	0.7561	&	\cellcolor{gray!60}0.7876	\\
		6	&	\cellcolor{gray!60}0.7710	&	{\bf 0.7372}	&	0.7392	&	\cellcolor{gray!60}0.7672	\\
		7	&	\cellcolor{gray!60}0.7579	&	{\bf 0.7254}	&	0.7266	&	\cellcolor{gray!60}0.7651	\\
		8	&	\cellcolor{gray!60}0.7499	&	0.7171	&	{\bf 0.7166}	&	\cellcolor{gray!60}0.7535	\\
		9	&	\cellcolor{gray!60}0.7409	&	0.7102	&	{\bf 0.7077}	&	\cellcolor{gray!60}0.7515	\\
		10	&	\cellcolor{gray!60}0.7366	&	0.7059	&	{\bf 0.7014}	&	\cellcolor{gray!60}0.7461	\\
		11	&	\cellcolor{gray!60}0.7307	&	0.7015	&	{\bf 0.6953}	&	\cellcolor{gray!60}0.7434	\\
		12	&	\cellcolor{gray!60}0.7266	&	0.6976	&	{\bf 0.6906}	&	\cellcolor{gray!60}0.7399	\\
		13	&	\cellcolor{gray!60}0.7218	&	0.6931	&	{\bf 0.6859}	&	\cellcolor{gray!60}0.738	\\
		14	&	\cellcolor{gray!60}0.7189	&	0.6901	&	{\bf 0.6818}	&	\cellcolor{gray!60}0.735	\\
		15	&	\cellcolor{gray!60}0.7153	&	0.6876	&	{\bf 0.6775}	&	\cellcolor{gray!60}0.7331	\\
		16	&	\cellcolor{gray!60}0.7133	&	0.6856	&	{\bf 0.6737}	&	\cellcolor{gray!60}0.7311	\\
		17	&	\cellcolor{gray!60}0.7101	&	0.6837	&	{\bf 0.6702}	&	\cellcolor{gray!60}0.7301	\\
		18	&	\cellcolor{gray!60}0.7084	&	0.6818	&	{\bf 0.6673}	&	\cellcolor{gray!60}0.7282	\\
		19	&	\cellcolor{gray!60}0.7059	&	0.6801	&	{\bf 0.6644}	&	\cellcolor{gray!60}0.7281	\\
		20	&	\cellcolor{gray!60}0.7043	&	0.6782	&	{\bf 0.6621}	&	\cellcolor{gray!60}0.7267	\\		
		\bottomrule						 
	\end{tabular}
\end{table}

Algo2 gives the best average error for training sets of size $\ell=1$ and for $3\leq \ell\leq 7$, while RF performs better for the other values of $\ell$. However, as the statistical tests indicate, for all sizes of the training sets except for $\ell=1$, there is no significant difference between the results produced by Algo2 and RF. For $\ell=1$ Algo2 is significantly better than the other 3 methods. Algo1 and SVR are statistically significantly worse than Algo2 for $\ell=1$ and also for $\ell>3$.  In summary, when there is only a single previous user rating, Algo2 is the preferable method. There is no significant difference between the four methods for $\ell=2$ and $3$. For training sets of size $\ell=5$ to 20, Algo2 and RF produce the best recommendations among all the methods, and the second-best methods are Algo1 and SVR.

\section{Conclusions and Future Work}\label{sec:5}
Nowadays, state of the art recommender systems are based on deep learning and graph embedding techniques, which are used in various platforms and domains \cite{Zhang22}. These systems need a lot of data and therefore are not suitable for small datasets where classical machine learning techniques may perform well. Also, the cold start problem is still a challenge, as even classical machine learning techniques need enough examples for training. 

To address the cold start problem, we have proposed in \cite{HKT21} a CB recommender system that builds user models on the basis of hypercube graphs. Limited experiments have shown that user preferences can be determined with a very limited number of ratings. In this paper, we have confirmed these findings on the basis of larger scale experiments. We have shown that  our method allows building user models with a very small number of ratings and therefore constitutes a particularly interesting approach to solve the cold start problem. In particular, experiments have shown that Algo1 and Algo2 outperform standard machine learning algorithms when the number of available ratings is at most 10, which often happens (see Figure \ref{fig:Figure1}). 

In all the experiments reported in Section \ref{sec:4}, the four competing methods which consistently produced the smallest average prediction error were Algo1, Algo2, RF and SVR. In the restaurants domain where we conducted a small-scale experiment, Algo1 gave the best predictions for small-sized training sets, with no statistical significant difference between the average prediction error obtained by the four methods. In the large-scale experiment conducted in the movie domain, Algo2 and RF were superior, with no statistical significant difference between them, where Algo1 and SVR got the second-best results. 

In summary, Tables \ref{tab:Table9} and \ref{tab:Table11} show that Algo2 produces statistically significantly smaller prediction errors (with very minor exceptions) than four out of the five tested machine learning methods. Note also that while these two tables contain results for 10 and 1000 users respectively, we see a similar trend that demonstrates the benefit of using Algo2 :
\begin{itemize}
	\setlength\itemsep{0pt}
\item SVR is statistically significantly worse than Algo2;
\item for small-sized training sets, Algo2 produces the smallest prediction error, but with no statistical significant difference when compared to RF;
\item for large-sized training sets, the results obtained by RF are the best but they are not statistically significantly different from the results obtained by Algo2. 
\end{itemize}

In order to get accurate predictions for any size of user ratings datasets, the methods can be combined. That is, we propose, as part of further research, to add to our model a hybrid mode which uses Algo1 and Algo2 for small size datasets, and when the number of user ratings is sufficient, the system switches to using RF or SVR for producing recommendations. However, the following should be considered when building a hybrid system. A major drawback of RF is that it can be slow because this method averages predictions obtained from a series of regression trees. To improve the precision of the predictions produced by RF, it is necessary to increase the number of trees which is computationally expensive \cite{Buskirk}. The SVR method, on the other hand, suffers from high training complexity.

A limitation of the research described in this paper is that the data for the experiments was extracted from the restaurant and movie domains only. Although the movie domain is commonly used for evaluating recommender systems algorithms \cite{Dau20,Wang20}, we would like to expand the experiments to other domains. 
Another limitation of this study is the use of a star scale in our models. However, the proposed algorithms can easily be adapted to deal with other rating scales. Other notable issues for future work are as follows.
\begin{itemize}
	\item {\em Privacy}. Preserving user privacy is a challenge that needs to be addressed since datasets may include sensitive and vulnerable information \cite{Himeur22}. The proposed recommendation models Algo1 and Algo2 allow preserving the confidential data of the users because the information necessary to identify the preferences of each user can be kept locally and does not require the knowledge of the ratings of other users. In future research we intend to implement Algo1 and Algo2 techniques on a laptop or a smartphone while using a mechanism for preserving user privacy, as proposed in \cite{Kuflik2009b}.
	\item {\em Resolving sets}. A subset $I'\subseteq I$ of items rated by a user $u\in U$ is a {\em resolving set} of the hypercube $Q_n$ if, given any two vertices $\boldsymbol{x},\boldsymbol{y}$ in $Q_n$, there is at least one item $i
	\in I'$ such that $d(\boldsymbol{x},\boldsymbol{v}^i)\neq d(\boldsymbol{y},\boldsymbol{v}^i)$ \cite{Belmonte}.
	In our context, this means that the knowledge of the Hamming distances between the vertex $\boldsymbol{w}^{u}$ associated with user $u$ and the vertices of a resolving set of $Q_n$ are sufficient to unequivocally determine the
	opinion of  $u$ on each attribute $a\in A$. As the resolving sets of hypercube graphs are of very small size \cite{Hertz20}, this approach can be used to solve the cold start problem. In future work, we will study how to encourage users of a recommender system to evaluate items that constitute a resolving set in order to know their preferences based on very few ratings.
	\end{itemize}

\parskip=-3pt
\bibliographystyle{acm}
\bibliography{ColdStartOctober112023}
\end{document}